\documentclass[epj]{svjour}
\usepackage{latexsym}
\usepackage{graphicx,color}
\usepackage{epsfig}
\usepackage{graphicx}
\usepackage{amsmath,amssymb}
\usepackage{color}

\begin{document}

\title{Entanglement in helium}

\author{Giuliano Benenti\inst{1,2} \and Stefano Siccardi\inst{3} 
\and Giuliano Strini\inst{4}}

\institute{                    
CNISM and Center for Nonlinear and Complex Systems,
Universit\`a degli Studi dell'Insubria, via Valleggio 11, 22100 Como, Italy
\and
Istituto Nazionale di Fisica Nucleare, Sezione di Milano,
via Celoria 16, 20133 Milano, Italy
\and
Department of Information Technologies, University of Milan,
via Bramante 65, 26013 Crema, Italy
\and
Department of Physics, University of Milan,
via Celoria 16, 20133 Milano, Italy
}

\abstract{
Using a configuration-interaction variational method, we accurately
compute the reduced, single-electron von
Neumann and linear entropy for several low-energy, singlet and triplet
eigenstates of helium atom.
We estimate the amount of electron-electron orbital
entanglement for such eigenstates and show that it
decays with energy.}


\authorrunning{G. Benenti, Si. Siccardi and G. Strini}
\titlerunning{Entanglement in helium}

\maketitle

\section{Introduction}
\label{sec:intro}

The role of entanglement as a resource in quantum 
communication and computation~\cite{qcbook,nielsen}
has stimulated many studies trying to unveil its 
fundamental aspects as well as to provide quantitative
entanglement measures~\cite{plenio,horodecki,fazio}.
More recently, the role of entanglement
attracts growing interest
in systems relevant for chemistry and biology.
For instance, the role of quantum coherence and
entanglement in atoms and molecules is investigated 
in laser-induced fragmentation experiments~\cite{dorner,becker}.
Moreover, entanglement is discussed in light-harvesting complexes,
governing the photosynthesis mechanism 
in proteins~\cite{engel,lee,mosheni,sarovar,huelga,buchleitner}
(for a critical view on the role of entanglement in
photosynthesis see, however, Ref.~\cite{briegel}).
In such instances, due to the complexity of the involved systems,
\emph{ab initio} treatments are computationally prohibitive
and one is forced to consider simplified (spin) models,
with parameters introduced phenomenologically.  

Besides computational problems for many-body quantum systems,
one must address the problem of measuring entanglement for
indistinguishable 
particles~\cite{schliemann,zanardi,wiseman,ghirardi,buchleitner2}. 
The main difficulty in quantifying 
entanglement is due to the symmetrization or
antisymmetrization of the wave function for bosons or fermions.
That is, to discriminate entanglement from correlations 
simply due to statistics of indistinguishable particles.
In spite of this difficulty, bipartite entanglement has
been investigated in a number of systems of physical interest,
including the entanglement dynamics 
of electron-electron scattering in low-dimensional semiconductor 
systems~\cite{bordone}, the changes of the electronic entanglement
during the dissociation process of diatomic molecules~\cite{esquivel},
and the entanglement of low-energy
eigenstates of helium atom~\cite{dehesa}. 

To evaluate entanglement in helium, Ref.~\cite{dehesa} 
used high-quality, state-of-the art 
Kinoshita-type wavefunctions~\cite{koga}, expressed in terms of
Hylleraas 
coordinates~\cite{hylleraas,koga,pekeris,kono,drake}, 
and then computed the 
purity of the reduced, one-electron density operator by means
of a twelve-dimensional Monte Carlo numerical integration. 
In the present paper, we compute the reduced, single-electron
von Neumann and linear entropy for several low-energy eigenstates of
helium by means of a simple 
configuration-interaction variational method. Our approach 
has several advantages. First of all, we do not need to evaluate 
multidimensional integrals: the reduced density matrix is 
obtained by purely algebraic methods. The reduced density matrix
can then be easily diagonalized and therefore we can access 
not only the linear entropy but also the von Neumann entropy. 
Finally, we express our variational, Slater-type basis, in terms of
(radial and angular) single-particle coordinates, and therefore
such Fock-state basis could be easily extended to many-body systems.
Despite the above mentioned, still unsolved conceptual 
difficulties in the 
definition of entanglement for indistinguishable particles,
we propose a way to evaluate the orbital 
entanglement for states close
to Fock states. From such a measure we conclude that the 
amount of entanglement exhibited by helium eigenstates 
$|1s,ns;{}^1S\rangle$ and $|1s,ns;{}^3S\rangle$
drops with energy. 

\newpage

\section{Entanglement of helium eigenstates}
\label{sec:enthelium}

The non-relativistic 
Hamiltonian of the helium atom reads, in atomic units,
\begin{equation}
H=\frac{1}{2}\,{\bf p}_1^2+\frac{1}{2}\,{\bf p}_2^2 -\frac{Z}{r_1}
-\frac{Z}{r_2}+\frac{1}{r_{12}},
\label{eq:hamiltonian}
\end{equation}
where $Z=2$ denotes the nuclear charge, ${\bf p}_i$ the momentum 
of electron $i$ ($i=1,2$), $r_i$ its distance
from the nucleus and $r_{12}$
the inter-electronic separation.

Since we are neglecting the spin-orbit interaction, we can consider
global wavefunctions $\Xi$ factorized into the product of a coordinate 
wavefunction $\Psi$ and a spin wavefunction $\chi$:
\begin{equation}
\Xi_{\sigma_1,\sigma_2}({\bf r}_1,{\bf r}_2)=
\Psi({\bf r}_1,{\bf r}_2)
\chi_{\sigma_1\sigma_2}.
\end{equation}
The overall state must be antisymmetric and therefore 
a measure~\cite{schliemann} of the amount of entanglement
$E(|\Xi\rangle)$ of the state $\Xi$ has been proposed 
in terms of the von Neumann entropy of the reduced density operator 
$R_1={\rm Tr}_2 (|\Xi\rangle\langle\Xi|)$ of one particle,
say particle 1, obtained after tracing the overall, two-body
density operator over the other particle:
\begin{equation}
E(|\Xi\rangle)=S(R_1)-1,
\label{eq:EXi}
\end{equation}
with the von Neumann entropy  
\begin{equation}
S(R_1)=-\sum_i \Lambda_i \log_2 \Lambda_i, 
\end{equation}
where $\{\Lambda_i\}$ are the eigenvalues of $R_1$.

However, with such definition a first problem arises.
When considering the triplet subspace, spanned by the spin states
$\chi_{\uparrow\uparrow}$, 
$\frac{1}{\sqrt{2}}(
\chi_{\uparrow\downarrow}+\chi_{\downarrow\uparrow})$, and 
$\chi_{\downarrow\downarrow}$, 
it is clear that we should consider 
the case $\frac{1}{\sqrt{2}}(
\chi_{\uparrow\downarrow}+\chi_{\downarrow\uparrow})$
separately from the cases  
$\chi_{\uparrow\uparrow}$ and $\chi_{\downarrow\downarrow}$.
Indeed, $\frac{1}{\sqrt{2}}
(\chi_{\uparrow\downarrow}+\chi_{\downarrow\uparrow})$ 
is a maximally entangled Bell state of the two spins while 
$\chi_{\uparrow\uparrow}$ and $\chi_{\downarrow\downarrow}$
are separable states. 
Therefore, the standard spectroscopic characterization in terms
of triplet and singlet states is no longer useful for the 
purposes of computing entanglement and one should study 
separately the entanglement properties of the states 
composing the triplet. In this context, we would like to
point out that, neglecting spin-spin interaction, the choice
of the basis states spanning the triplet state is completely
arbitrary and that the above discussed 
$\{\chi_{\uparrow\uparrow},
\frac{1}{\sqrt{2}}(
\chi_{\uparrow\downarrow}+\chi_{\downarrow\uparrow}),
\chi_{\downarrow\downarrow}\}$
is only one in between the infinite possible choices. 

To avoid this ambiguity, in this paper we compute 
the entanglement for the orbital part $\Psi$ only of the  
wavefunction.
Since $\Psi$ is antisymmetric for spins in the triplet subspace,
we can measure the amount of entanglement 
$E(|\Psi\rangle)$ of $\Psi$ as follows:
\begin{equation}
E(|\Psi\rangle)=S(\rho_1)-1,
\label{eq:entfermions}
\end{equation}
where 
\begin{equation}
S(\rho_1)=-\sum_i \lambda_i \log_2 \lambda_i 
\end{equation}
is the von Neumann entropy of the reduced density operator 
$\rho_1={\rm Tr}_2 (|\Psi\rangle\langle\Psi|)$,
and $\{\lambda_i\}$ are the eigenvalues of $\rho_1$.
 
When the spin part of the wavefunction is in the singlet state
$\chi_S=\frac{1}{\sqrt{2}}(
\chi_{\uparrow\downarrow}-\chi_{\downarrow\uparrow})$,
the orbital part is necessarily symmetric and this causes
an additional, open problem in the quantification of entanglement.
Indeed in this case the reduced von Neumann entropy alone is
not sufficient to discriminate between entangled and separable 
states~\cite{ghirardi}. 
The core of the problem is the fact that we can have separable states
with either $S(\rho_1)=0$ or $S(\rho_1)=1$. The first instance 
corresponds to basis states of the kind $\Psi_{ii}({\bf r}_1,{\bf r}_2)=
\phi_i({\bf r_1})\phi_i({\bf r}_2)$, the second to basis states like
$\Psi_{ij}({\bf r}_1,{\bf r}_2)=
\frac{1}{\sqrt{2}}[\phi_i({\bf r_1})\phi_j({\bf r}_2)+
\phi_j({\bf r_1})\phi_i({\bf r}_2)]$, with $i\ne j$.
On the other hand, even the quantification of entanglement
of the global, antisymmetric wavefunction by means of 
Eq.~(\ref{eq:EXi}) poses a problem. Indeed, 
as the von Neummann entropy is additive for tensor products,
for the state
$\Xi=\Psi_{ii}\chi_S$
Eq.~(\ref{eq:EXi}) gives 
$E(|\Xi\rangle)=0$, while 
for the state $\Xi=\Psi_{ij}\chi_S$  ($i\ne j$) we have
$E(|\Xi\rangle)=1$. Even though measure (\ref{eq:EXi}) 
gives different results, the amount of entanglement 
in both cases is the same, since the orbital wavefunctions
$\Psi_{ii}$ and $\Psi_{ij}$ are both separable: the reduced
density matrices for these two states have different entropies
only due to the symmetrization of the state $\Psi_{ij}$.  

We will not address in this paper the unsolved 
problem of quantification
of entanglement for a generic state $\Psi$. 
On the other hand, since from our calculations it turns out that 
for each low-energy helium eigenstate the reduced density operator for 
the orbital part is rather weakly perturbed with respect to 
one of the two above non-entangled cases, we expect that an
approximate quantification
of entanglement is provided
by the distance between the von Neumann entropy $S(\rho_1)$ of $\rho_1$ 
and the von Neumann entropy $S(\rho_1^{(0)})$
($S(\rho_1^{(0)})=0$ or $S(\rho_1^{(0)})=1$) of the 
reduced density operator $\rho_1^{(0)}$ for the corresponding 
non-interacting, non-entangled state:
\begin{equation}
E(|\Psi\rangle)=|S(\rho_1)-S(\rho_1^{(0)})|.
\label{eq:entbosons}
\end{equation}
We do not intend to propose
an entanglement measure in a rigorous sense; that is not one 
satisfying all the requirements
listed e.g. in \cite{plenio};
in particular we
are not going to show that it is not increasing under 
Local General Measurements and
Classical Communication. Our aim is just to have some provisional 
heuristics that could
put some order in the data.
We expect such quantification to be in general 
meaningful only in the regime of weak interactions, such that 
$|S(\rho_1)-S(\rho_1^{(0)})|\ll 1$.
Note, however, that for antisymmetric orbital wave functions this definition
reduces to (\ref{eq:entfermions}) and therefore could be applied 
in this case also for strong interactions.

\section{Method}
\label{sec:method}

We compute with high accuracy 
the lowest energy eigenstates of helium 
by means of a variational method,
the configuration-interaction method (see, for instance,
Ref.~\cite{fulde}). Orthonormal basis functions are provided by
\begin{equation}
\begin{array}{c}
\Phi_{n_1,l_1,m_1;n_2,l_2m_2}({\bf r}_1,{\bf r}_2)
\\
\\
=F_{n_1,l_1;n_2,l_2}(r_1,r_2) 
Y_{l_1m_1}(\Omega_1) Y_{l_2 m_2}(\Omega_2),
\end{array}
\end{equation}
where $Y_{l_i m_i}$ are spherical harmonics, with 
$\Omega_i$ solid angle for particle $i$ and the radial 
functions $F_{n_1,l_1;n_2,l_2}(r_1,r_2)$ are obtained 
after orthonormalizing the Slater-type orbitals
\begin{equation}
R_{n l} (r)= r^{n+l-1} \exp(-\xi_{n,l} r),
\end{equation}
with $\xi_{n,l}$ variational parameters,
and properly symmetrizing the products of the obtained 
one-particle radial wavefunctions $f_{nl}(r)$. 
That is, if the spin wavefunction is symmetric, 
$F$ must be antisymmetric,
\begin{equation}
\begin{array}{c}
{\displaystyle
F_{n_1,l_1;n_2,l_2}(r_1,r_2)=
\frac{1}{\sqrt{2}} 
[f_{n_1 l_1}(r_1) f_{n_2 l_2}(r_2)
}
\\
\\
-f_{n_2 l_2}(r_1) f_{n_1 l_1}(r_2)];
\end{array}
\end{equation}
if the spin wavefunction is antisymmetric,
$F$ must be symmetric:
\begin{equation}
\begin{array}{c}
{\displaystyle
F_{n_1,l_1;n_2,l_2}(r_1,r_2)=
\frac{1}{\sqrt{2}} 
[f_{n_1 l_1}(r_1) f_{n_2 l_2}(r_2)
}
\\
\\
+f_{n_2 l_2}(r_1) f_{n_1 l_1}(r_2)]
\end{array}
\end{equation}
if $(n_1,l_1)\ne (n_2,l_2)$, 
\begin{equation}
F_{n_1,l_1;n_1,l_1}(r_1,r_2)=f_{n_1 l_1}(r_1) f_{n_1 l_1}(r_2)
\end{equation}
otherwise.

We then compute the reduced (single-electron) density matrix 
\begin{equation}
\rho_1({\bf r}_1,{\bf r}_1^\prime)=
\int d{\bf r}_2 \Psi({\bf r}_1,{\bf r}_2) 
\Psi^\star({\bf r}_1^\prime,{\bf r}_2),
\label{eq:integrals}
\end{equation}
with 
\begin{equation}
\Psi({\bf r}_1,{\bf r}_2)=\sum_{I_1,I_2} c_{I_1,I_2} 
\Phi_{I_1,I_2} ({\bf r}_1,{\bf r}_2),
\end{equation}
with the multi-indexes $I_1\equiv(n_1,l_1,m_1)$ and
$I_2\equiv(n_2,l_2,m_2)$.
Since the expansion is done over an orthonormal basis
the reduced density matrix on that basis is simply 
given by a partial trace over the second particle of the
overall density matrix:
$(\rho_1)_{I_1,I_1^\prime}=\sum_{I_2}
\rho_{I_1,I_2;I_1^\prime,I_2}$,
where $\rho_{I_1,I_2;I_1^\prime,I_2^\prime}
=c_{I_1I_2} c_{I_1^\prime I_2^\prime}^\star$.
We point out a major advantage of the 
configuration-interaction method and the use of 
orthonormal orbitals: the reduced density matrix is 
obtained by purely algebraic methods, without 
numerical computation of the multi-dimensional 
integrals of Eq.~(\ref{eq:integrals}).
The reduced density matrix can then be easily diagonalized and in
this paper we will study the entanglement properties
of helium by means of the eigenvalues $\lambda_k$ 
of $\rho_1$.
We will limit our investigation to the ground state
and singly excited eigenstates of helium. While doubly excited 
states can also be addressed by our method, their investigation
requires, as well known in the 
literature~\cite{cormier,delande,scrinzi,buchleitner3,madronero},
much larger basis dimensions.

\section{Numerical results}
\label{sec:numerics}

In the following, we will present data only for total orbital 
momentum quantum number $L=0$, thus implying $l_1=l_2\equiv l$
and $m_1=-\,m_2$.

\begin{figure}[htp]
\includegraphics[angle=0.0, width=8cm]{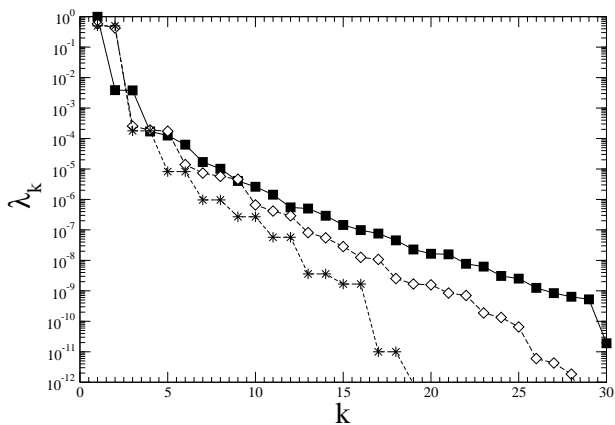}
\caption{Eigenvalue spectrum of the reduced density matrix 
$\rho_1$, for the states 
$|(1s)^2;{}^1S\rangle$ (squares),
$|1s,2s;{}^1S\rangle$ (diamonds), and
$|1s,2s;{}^3S\rangle$ (stars). In these computations 
$n_{\rm max}=10$ and $l_{\rm max}=2$.}
\label{fig:spectrum}
\end{figure}

We first discuss convergence of our method, as a function
of the number $n_{\rm max}(l)$ of radial wavefunctions for a given $l$ and
as a function of the cut-off $l_{\rm max}$ on $l$.
For the low-energy states discussed below, we found that 
$l_{\rm max}=2$ (S, P, and D shells) and $n_{\rm max}\approx 10-20$
(from $n_{\rm max}=10$ for the ground state up to 
$n_{\rm max}=20$ for the highest excited states reported below)
are sufficient to reproduce helium eigenergies with at least
four significant digits (as deduced from comparison of our
results with those of Refs.~\cite{koga,pekeris,kono,drake}) 
and reduced von Neumann entropy 
$S(\rho_1)$, estimating at least 
two-three significant digits. 
To illustrate the convergence of our method, we provide 
in Table~\ref{table:convergence} the obtained values of 
$S(\rho_1)$ of the ground state of helium 
for different values of the cut-offs 
$l_{\rm max}$ and $n_{\rm max}$
(we take the same $n_{\rm max}$ for all 
values of $l$).
We also show in Fig.~\ref{fig:spectrum} the spectrum 
of the reduced density matrix $\rho_1$, that is, the eigenvalues 
$\lambda_k$ versus $k$, for a few exemplary cases:
the ground state $|(1s)^2;{}^1S\rangle$ and the lowest energy 
excited states $|1s,2s;{}^1S\rangle$ and $|1s,2s;{}^3S\rangle$. 
We note that in the case of the ground state there is a single 
dominant eigenvalue, $\lambda_1\approx 0.992$, while for the 
singlet excited state we have two dominant eigenvalues,
$\lambda_1\approx 0.577$ and $\lambda_2\approx 0.422$. 
Finally, for the triplet states the orbital wave function is
antisymmetric and therefore the eigenvalues are doubly degenerate. 
For the states $|1s,2s;{}^3S\rangle$ we have 
$\lambda_1=\lambda_2\approx 0.4998$. The different features of the 
spectrum in the above described cases are consistent with the 
different values of the reduced von Neumann and linear entropies, 
in spite of the fact that we are always close to the 
non-interacing limit where the eigenstates are single
Slater permanents or determinants. 

\begin{table}[h]
\begin{center}
\begin{tabular}{|l|c|c|c|c|}
\hline
& $l_{\rm max}=0$ & $l_{\rm max}=1$ & $l_{\rm max}=2$ 
& $l_{\rm max}=3$\\ 
\hline
 $n_{\rm max}=5$  & 0.04131 & 0.07772 & 0.07844 & 0.07833 \\
 $n_{\rm max}=6$  & 0.04133 & 0.07776 & 0.07848 & 0.07837 \\
 $n_{\rm max}=10$ & 0.04134 & 0.07777 & 0.07849 & 0.07839 \\
 $n_{\rm max}=11$ & 0.04134 & 0.07777 & 0.07849 & 0.07839 \\
\hline
\end{tabular}
\end{center}
\caption{Reduced von Neumann entropy of the ground state of
helium, computed with different cut-off values 
in the basis of Slater-type orbitals.}
\label{table:convergence}
\end{table}

The reduced von Neumann entropy $S(\rho_1)$, as well as the linearized
entropy $S_L(\rho_1)= 1- {\rm Tr}(\rho_1^2)$ 
often used in the literature, are shown in Table~\ref{table:singlet} 
and in Table~\ref{table:triplet} for several low-energy 
singlet and triplet eigenstates,
respectively. 

\begin{table}[ht]
\begin{center}
\begin{tabular}{|c|c|c|}
\hline
\hbox{State} & $S(\rho_1)$ & $S_L(\rho_1)$ \\ 
\hline
$|(1s)^2;{}^1S\rangle$ & 0.0785 & 0.01606 \\
$|1s,2s;{}^1S\rangle$   & 0.991099 & 0.48871 \\  
$|1s,3s;{}^1S\rangle$   & 0.998513 & 0.49724 \\  
$|1s,4s;{}^1S\rangle$   & 0.999577 & 0.49892 \\  
$|1s,5s;{}^1S\rangle$   & 0.999838 & 0.499465 \\  
$|1s,6s;{}^1S\rangle$   & 0.999923 & 0.499665 \\  
$|1s,7s;{}^1S\rangle$   & 0.999961 & 0.499777 \\  
\hline
\end{tabular}
\end{center}
\caption{Reduced von Neumann and linear entropies for the 
lowest energy singlet eigenstates of helium.}
\label{table:singlet}
\end{table}

\begin{table}[ht]
\begin{center}
\begin{tabular}{|c|c|c|}
\hline
\hbox{State} & $S(\rho_1)$ & $S_L(\rho_1)$ \\ 
\hline
$|1s,2s;{}^3S\rangle$   & 1.00494  & 0.500378 \\  
$|1s,3s;{}^3S\rangle$   & 1.00114  & 0.5000736 \\  
$|1s,4s;{}^3S\rangle$   & 1.000453 & 0.5000267 \\  
$|1s,5s;{}^3S\rangle$   & 1.000229 & 0.5000127 \\  
$|1s,6s;{}^3S\rangle$   & 1.000133 & 0.5000070 \\  
$|1s,7s;{}^3S\rangle$   & 1.000091 & 0.5000047 \\  
\hline
\end{tabular}
\end{center}
\caption{Same as in Table~\ref{table:singlet}, but for 
the lowest energy triplet eigenstates of helium.}
\label{table:triplet}
\end{table}

\begin{figure}[htp]
\includegraphics[angle=0.0, width=8cm]{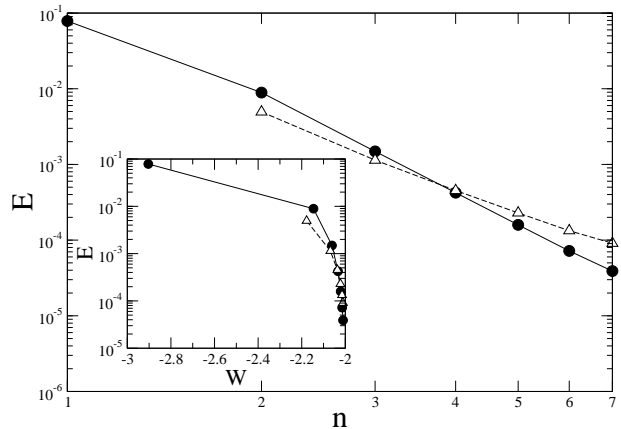}
\caption{Entanglement of the singlet states 
$|1s,ns;{}^1S\rangle$ (circles) and 
of the triplet states $|1s,ns;{}^3S\rangle$ (triangles)
as a function of $n$ (main plot, with logarithmic scale 
for both axes) and of the
energy $W$ of the eigenstates, measured in hartrees 
(inset, logarithmic scale only for the $E$-axis).}
\label{fig:entanglement}
\end{figure}

Since the obtained values of the von Neumann entropy are 
very close to those expected for Fock states, which are
separable, the entanglement content is weak and can 
be estimated by means of Eq.~(\ref{eq:entbosons}). 
The obtained results are shown in Fig.~\ref{fig:entanglement}
as a function of the 
state number $n$, 
for both singlet states $|1s,ns;{}^1S\rangle$ 
and triplet states $|1s,ns;{}^3S\rangle$.
Note that data, with the exception of the ground state
value of entanglement, are consistent with a power law decay 
of entanglement with $n$. From a power-law fit we obtained 
$E(n)=0.19 \, n^{-4.41}$ for singlet states at $n\ge 2$ 
and $E(n)= 0.040 \, n^{-3.19}$ for triplet states.
The same entanglement data are shown as a function of
energy in the inset of Fig.~\ref{fig:entanglement}. 
It can be clearly seen that the entanglement content 
drops with energy. This result is rather intuitive in 
that for states $|1s,ns;{}^1 S\rangle$ and 
$|1s,ns;{}^3 S\rangle$ the wave functions corresponding to 
the states $1s$ and $ns$ are localized farther apart for larger 
$n$. Therefore, electron-electron interactions become
weaker (and entanglement smaller) when $n$ increases. 

We note that, due to a different definition of 
entanglement, the energy-dependence obtained in 
our results contrasts the one obtained
in Ref.~\cite{dehesa}.   
Our present approach differs from the one in 
Ref.~\cite{dehesa} 
because we focus exclusively on the entanglement associated with the
spatial part of the two-electron wave function, while Ref.~\cite{dehesa}
considers the entanglement associated with the global two-electron state,
taking spin explicitly into account.
In that paper, the amount of entanglement of 
state $\Xi=\Psi\chi$ is defined as ${\cal E}(\Xi)=
2\left[S_L(R_1)-\frac{1}{2}\right]=1-2{\rm Tr}(R_1^2)$,
where ${\rm Tr}(R_1^2)={\rm Tr}(\rho_1^2){\rm Tr} (\rho_{s1}^2)$,
with $\rho_{s1}$ reduced density operator for the spin
wavefunction. 
Such definition causes problems. 
For instance, for the singlet eigenstates
we have ${\rm Tr} (\rho_{s1}^2)=\frac{1}{2}$, and therefore
${\cal E}(\Xi)=1-{\rm Tr}(\rho_1^2)=S_L(\rho_1)$, which
is a growing function of energy
(see Table~\ref{table:singlet}).
On the other hand,
this definition does not take into account the fact that 
$S_L(\rho_1)$ for the state 
$|(1s)^2;{}^1S\rangle$ is much smaller than for the states
$|1s,ns;{}^1S\rangle$, with $n>1$, not due to 
a smaller entanglement content but as a consequence of 
the symmetrization of the orbital part of the wavefunction.
Indeed, as discussed at the end of  
the Sec.~\ref{sec:enthelium}, for the state
$|(1s)^2;{}^1S\rangle$ we have 
$S(\rho_1^{(0)})=0$ and $S_L(\rho_1^{(0)})=0$, while 
for the states $|1s,ns;{}^1S\rangle$ we have
$S(\rho_1^{(0)})=1$ and $S_L(\rho_1^{(0)})=\frac{1}{2}$,
with $\rho_1^{(0)}$ orbital density operator when
electron-electron interaction is neglected.
Since state $\rho_1^{(0)}$ is not entangled and 
$S_L(\rho_1)\approx S_L(\rho_1^{(0)}$), the entanglement
content of state $\Psi$ cannot be properly estimated
by $S_L(\rho_1)$, but one should rather consider the 
difference $|S_L(\rho_1)-S_L(\rho_1^{(0)})|$ or, as 
we have proposed in this paper, $|S(\rho_1)-S(\rho_1^{(0)})|$. 

We also point out that a measure of the degree of 
quantum correlations in multipartite systems was proposed 
in Ref.~\cite{eberly}. In that paper, the degree of 
correlations of a $N$-particle bosonic of fermionic system 
was estimated by means of the inverse participation
ratio of the eigenvalues $\lambda_k$ of the reduced 
single-particle density operator, multiplied by a factor 
$N$ in the case of fermions to take into account the 
$N$-fold degeneracy of eigenvalues. However, such 
definition does not take into account the fact that, for
symmetric wave functions, the degree of degeneracy of the 
eigenvalues $\lambda_k$ is not unique, see the above described
difference between the states 
$|(1s)^2;{}^1S\rangle$ and
$|1s,ns;{}^1S\rangle$. 

\section{Conclusions}
\label{sec:conc}

In this paper, we have computed the reduced von Neumann entropy
for several low-energy (singlet and triplet) eigenstates of
helium. The von Neumann entropy has then been used to 
estimate the amount of entanglement of such states,
showing that the entanglement of 
the states $|1s,ns;{}^1S\rangle$ and 
$|1s,ns;{}^3S\rangle$ decays with $n$, that is,
with energy. This result is in agreement with the 
intuition, suggesting that when the electronic wavefunctions  
are localized far apart, electron-electron interactions
are weak and therefore the entanglement is expected to be
small. 
Our results can be seen as one of the first steps towards
a ``spectroscopy of entanglement'' for atomic and molecular
systems. While in this quest helium atom constitutes the simplest 
example, the variational scheme used in this paper may be 
extended to more complex systems, and the obtained numerical 
results could be used to test the validity,
with respect to the calculation of the reduced 
von Neumann or linear entropy, of perturbative 
approaches or of simplified phenomenological models.
Finally, we point out that, while our bipartite 
entanglement measure can be applied to wavefunctions
close to separable Fock states, quantification of 
entanglement for generic states of indistinguishable 
particles remains an interesting open problem.

\section*{Acknowledgements}

We thank Jes\'us Dehesa for very useful discussions.

\vspace{3pc}


\end{document}